\documentstyle[aps,12pt]{revtex}

\begin{document}
\title{Ballistic transport in one-dimensional loops with Rashba and Dresselhaus
spin-orbit coupling }
\author{V. Marigliano Ramaglia$^1$, V. Cataudella$^1$, G. De Filippis$^1$ and C.A.
Perroni$^2$}
\address{$^1${\sl Coherentia CNR-INFM and Dipartimento di Scienze Fisiche,}\\
{\sl Universit\`{a} degli Studi di Napoli ``Federico II'', 80126 Napoli,}\\
Italy\\
$^2${\it I}{\sl nstitute of Solid State Research (IFF), Research Center}\\
J\"{u}lich, D-52425 J\"{u}lich, Germany.}
\date{\today}
\maketitle

\begin{abstract}
We discuss the combined effect of Rashba and Dresselhaus spin-orbit
interactions in polygonal loops formed by quantum wires, when the electron
are injected in a node and collected at the opposite one. The conditions
that allow perfect localization are found. Furthermore, we investigate the
suppression of the Al'tshuler--Aronov--Spivak oscillations that appear, in
presence of a magnetic flux, when the electrons are injected and collected
at the same node. Finally, we point out that a recent realization of a
ballistic spin interferometer can be used to obtain a reliable estimate of
the magnitude ratio of the two spin-orbit interactions.\bigskip

PACS numbers: 71.70.Ej,73.23.Ad
\end{abstract}

\section{Introduction}

The main goal of the spintronics is the manipulation of spins in
semiconductor nanostructures. To this aim a large number of devices
exploiting spin-orbit (SO) interactions\cite
{Awschalom,Datta,Marigliano,Koga1,Pala,Governale1} has been proposed. One of
these interactions, known as ``Rashba Effect'' \cite{Rashba}, appears at the
interface of semiconductors lacking of structural inversion symmetry and its
magnitude can be controlled by an applied gate voltage. The devices based on
this effect use the quantum interference, due to the spin precession, beween
different paths. Among the others we remind the ballistic spin interferometer%
\cite{Koga2}, in which a square loop is followed along a self-intersecting
trajectory in clockwise and anticlockwise way, that, recently, has been used
to demonstrate experimentally the occurence of the spin precession
interference phenomenon\cite{Koga3}. In particular, the suppression of the
Al'tshuler--Aronov--Spivak (AAS) oscillations\cite{Al'tshuler} allows the
measurement of the magnitude of the Rashba interaction, and Koga et al. \cite
{Koga3} have obtained values in agree with theoretical estimates and with
the Weak Antilocalization Analysis. Besides it has been shown that the
Rashba effect is also able to induce localization effects in quantum networks%
\cite{Frustaglia,Bercioux1,Bercioux2}.

The inversion asymmetry in the bulk semiconductor gives rise to
spin-dependent bulk band structure. At the surface this SO interaction,
known as ``Dresselhaus term'' \cite{Dresselhaus}, adds to the Rashba term.
Recent measurements based on the spin-galvanic effect provided the ratio
between magnitude of Rashba and Dresselhaus terms. This ratio can reach
values as large as 2.14$\pm $0.25 in InAs quantum well\cite{Ganichev1}. The
Rashba term is in general dominant but the Dresselhaus interaction can have
observable effects.

In a quantum wire the two SO couplings yield together a spin precession
depending on the angular position of the wire\cite{Ganichev2}. In the
experiments by Ganichev et al.\cite{Ganichev1}, a circularly polarized light
produces a spin galvanic current whose intensity exhibits an angular
dependence that allows the measure of the ratio between the SO couplings.
Schliemann et al. \cite{Schliemann} have proposed a spin-field-effect
transistor in which the presence of the two SO couplings with equal
magnitudes can give polarized currents whose spin does not depend on the
momentum. In such a way the spin-independent scattering processes become
ineffective in the particular direction in which the spin precession is
suppressed.

In this paper we study the interference effects in one--dimensional loops
due to spin precession when both the two SO interactions are present. The
paper is organized in the following way. In order to be self-contained in
the section II we recall a number of already known results\cite
{Ganichev1,Schliemann} that will be used to describe the spin precession in
a quantum wire under the two SO couplings\cite{vanVee}. In the section III
we show how the localization in a polygonal loop can be achieved. We
emphasize that for a diamond square loop with the diagonal oriented in [010]
crystallografic direction there is a periodic set of values of the SO
strengths that gives perfect localization, i.e. the transmission coefficient
vanishes. Rotating the diamond square loop the localization is lost. We also
show that for particular rhombic and exagonal loops the transmission
vanishes only at specific values of the SO strengths. In the section IV we
consider what happens when a magnetic flux threads the loop, i.e. we analyse
a ballistic spin interferometer with both the SO couplings. Particular
attention will be paid to the suppression of the AAS oscillations that
appear when the input and the ouput node coincide. We will see how the SO
magnitudes ratio shifts the values of the Rashba SO strength at which the
transmission becomes independent on the magnetic flux. Finally we prove that
the Aharonov-Bohm (AB) oscillatons appearing when we inject and collect the
current in opposite nodes, can be also modulated varying the two SO
couplings. The section V is dedicated to some concluding remarks.

\section{Spin precession due to Rashba and Dresselhaus coupling}

\subsection{The spin-orbit couplings in a two dimensional electron gas}

In order to set the notation let us remind the eigenstates and the energy
eigenvalues of an electron confined in the $x-z$ plane and subjected to both
Rashba and Dresselhaus spin-orbit interaction\cite{Schliemann}. The
hamiltonian takes the form 
\begin{equation}
H=\frac{\hbar ^2}{2m}\left( p_x^2+p_z^2\right) +H_R+H_D  \label{eq1}
\end{equation}
where 
\begin{equation}
H_R=\frac \alpha \hbar \left( \sigma _zp_x-\sigma _xp_z\right)  \label{eq2}
\end{equation}
and 
\begin{equation}
H_D=\frac \beta \hbar \left( \sigma _zp_z-\sigma _xp_x\right) ,  \label{eq3}
\end{equation}
are the Rashba and the Dresselhaus interactions, respectively. We have
choosen the $x-$axis and $z-$axis in [010] and [100] crystallographic
directions, respectively. It is easy to check that 
\begin{equation}
\psi _{\vec{k}\,\,,\,\,\pm }\left( x,z\right) =\exp \left[
i(k_xx+k_zz)\right] \left| 
\begin{array}{c}
\cos \nu _{\pm } \\ 
\sin \nu _{\pm }
\end{array}
\right|  \label{eq3a}
\end{equation}
are eigenfunctions of (\ref{eq1}) with eigenvalues given by 
\begin{equation}
E_{\pm }=\frac{\hbar ^2}{2m}k^{\,2}\pm \sqrt{\left( \alpha ^2+\beta
^2\right) k^2+4\alpha \beta k_xk_z}  \label{eq4}
\end{equation}
where $k=\sqrt{k_x^2+k_z^2}$ ($k_x=k\cos \theta \,,\,k_z=k\sin \theta $) is
the modulus of the momentum in $x-z$ plane. In eq.(\ref{eq3a}) we have
defined 
\begin{equation}
\nu _{\pm }=\arctan \frac{k_0\cos \theta +k_1\sin \theta \mp k_{so}\left(
\theta \right) }{k_0\sin \theta +k_1\cos \theta }  \label{eq4a}
\end{equation}
where 
\[
\,\,\,\,k_{so}\left( \theta \right) =\sqrt{k_0^2+k_1^2+2k_0k_1\sin 2\theta }%
\,\,\,\text{with }k_0=\frac{m\alpha }{\hbar ^2}\,\,\,\text{and}\,\,\,k_1=%
\frac{m\beta }{\hbar ^2}\,. 
\]

We note that there are two values of $k$ corresponding to the same energy $E=%
\frac{\hbar ^2}{2m}\xi ^2$ and they are given by 
\begin{equation}
k=k_{\pm }=\sqrt{\xi ^2+k_{so}^2}\mp k_{so}  \label{eq5}
\end{equation}
with the corresponding energy that can be rewritten as 
\[
E_{\pm }=E=\frac{\hbar ^2}{2m}\left( k_{\pm }^2\pm 2k_{\pm }k_{so}\left(
\theta \right) \right) . 
\]
The spinors $\chi _{\pm }$ of the two degenerate modes are orthogonal each
other, being 
\[
\nu _{-}=\frac \pi 2+\nu _{+}\,\,, 
\]
therefore we have 
\[
\chi _{+}=\left| 
\begin{array}{l}
\cos \nu _{+} \\ 
\sin \nu _{+}
\end{array}
\right| \,\,\,\,\text{and}\,\,\,\,\chi _{-}=\left| 
\begin{array}{l}
-\sin \nu _{+} \\ 
\cos \nu _{+}
\end{array}
\right| . 
\]

It is worth to note that with the only Rashba interaction $\left(
k_1=0\right) $ we have 
\[
\nu _{+}=-\frac \theta 2. 
\]
We remind that the Rashba SO interaction can be viewed as a magnetic field
parallel to the plane and orthogonal to the wavevector $\vec{k}$ that
orientates the spin along the direction perpendicular to the wave vector\cite
{Marigliano1}. In particular when the mode $\left( -\right) $ propagates in $%
x-$direction the spinor $\chi _{-}$ $=\left| 
\begin{array}{c}
0 \\ 
1
\end{array}
\right| $ is in the spin down state along $z-$direction. On the other hand
with only Dresselhaus interaction $\left( k_0=0\right) $ we have 
\[
\nu _{+}=-\frac \pi 4+\frac \theta 2 
\]
and the SO magnetic field is opposite to $\vec{k}$. Now, when the mode $%
\left( -\right) $ propagates in $x-$direction, $\chi _{-}$ $=$ $\frac 1{%
\sqrt{2}}\left| 
\begin{array}{c}
1 \\ 
1
\end{array}
\right| $ and the spin is oriented along $x-$axis. When both the SO
interactions are present the effective SO magnetic field, parallel to the
plane, fixes the spin direction according to eq.(\ref{eq4a}).

\subsection{Spin precession in a quantum wire due to the spin-orbit
interactions}

Let us assume that an electron moves in a one-dimensional (1D) ballistic
quantum wire along an arbitrary $\theta -$direction and subjected to
spin-orbit interactions. Moreover, we neglect the subband hybridization,
induced by the spin-orbit coupling, assuming that the quantum wire is a
truly 1D system because the spin--precession length $\pi /k_{SO}$ is much
larger than the wire width\cite{nota1}. Within our approximation the
spin-splitted bands have the orbital parts given by $e^{ik_{\pm }r}$ ( $r$
is the coordinate along $\theta -$direction).

In order to calculate the spin-orbit precession along the wire direction we
proceed in the following way (see also van Veehuizen et al.\cite{vanVee}).
First of all we project an arbitrary input spin state in $r=0$ 
\[
|\psi \left( 0\right) \rangle =\left| 
\begin{array}{c}
a \\ 
b
\end{array}
\right| 
\]
on $\chi _{\pm }$ spinors, obtaining 
\[
\left\langle \chi _{+}|\psi \left( 0\right) \right\rangle
=ac_{+}+bs_{+}\,\,\,;\,\,\,\left\langle \chi _{-}|\psi \left( 0\right)
\right\rangle \,=-as_{+}+bc_{+}\,\,, 
\]
where 
\[
c_{+}=\cos \nu _{+}\,\,\,\,\,\text{and}\,\,\,\,\,s_{+}=\sin \nu _{+}. 
\]
Then, after a displacement $L$ along $\theta -$direction, the electron will
be in the state $|\psi \left( L\right) \rangle $ given by 
\[
|\psi \left( L\right) \rangle =e^{ik_{+}L}\left( ac_{+}+bs_{+}\right) |\chi
_{+}\rangle +e^{ik_{-}L}\left( -as_{+}+bc_{+}\right) |\chi _{-}\rangle . 
\]
It easy to show that $|\psi \left( L\right) \rangle $ can be written in
terms of the spin initial state $|\psi \left( 0\right) \rangle :$ 
\begin{equation}
|\psi \left( L\right) \rangle =\left| 
\begin{array}{cc}
c_{+}^2e^{ik_{+}L}+s_{+}^2e^{ik_{-}L} & s_{+}c_{+}\left(
e^{ik_{+}L}-e^{ik_{-}L}\right) \\ 
s_{+}c_{+}\left( e^{ik_{+}L}-e^{ik_{-}L}\right) & 
s_{+}^2e^{ik_{+}L}+c_{+}^2e^{ik_{-}L}
\end{array}
\right| \cdot \left| 
\begin{array}{c}
a \\ 
b
\end{array}
\right| .  \label{eq5a}
\end{equation}
Introducing the spin operator ${\bf R}_{SO}$%
\begin{equation}
{\bf R}_{SO}=\left| 
\begin{array}{cc}
\cos k_{so}L-i\cos 2\nu _{+}\sin k_{so}L & -i\sin k_{so}L\sin 2\nu _{+} \\ 
-i\sin k_{so}L\sin 2\nu _{+} & \cos k_{so}L+i\cos 2\nu _{+}\sin k_{so}L
\end{array}
\right| \,\,,  \label{eq8}
\end{equation}
the eq.(\ref{eq5a}) can be also written as 
\begin{equation}
|\psi \left( L\right) \rangle ={\bf R}_{SO}\,e^{i\sqrt{\xi ^2+k_{so}^2}%
L}|\psi \left( 0\right) \rangle .  \label{eq7}
\end{equation}
In the following we assume that $\xi ^2\gg k_{so}^2\left( \theta \right) $
because, in the realistic systems, the strength of SO, $k_{SO}T$, ranges
from 0.01$\xi $ to 0.05$\xi $, where $\xi $ is the Fermi wavevector\cite
{urnitta}. Therefore we take the orbital part with $k_{\pm }\cong \xi \mp
k_{so}$, neglecting terms of the second order in $\xi /k_{so}$ . Then, only
the spin operator ${\bf R}_{SO}$ depends on the angular position of the wire
while the dynamical phase factor become equal to $\exp \left( i\xi L\right)
. $ The matrix ${\bf R}_{SO}$, actually, describes a geometrical rotation in
the $\frac 12\,$spin space around the unitary vector 
\[
\vec{u}=\left( \sin 2\nu _{+},0,\cos 2\nu _{+}\right) 
\]
of the angle $2k_{so}L.$ In fact $R_{SO}$ is the representation of the
rotation operator\cite{Cohen} 
\begin{equation}
{\bf R}_{SO}=\exp \left( -i\,k_{so}L\,\,{\bf \vec{\sigma}}\cdot \vec{u}%
\right) =\cos k_{so}L\otimes {\bf 1}-i\sin k_{so}L\otimes \;{\bf \vec{\sigma}%
}\cdot \vec{u}^{}  \label{eq9}
\end{equation}
where ${\bf 1}$ is the unit matrix and ${\bf \vec{\sigma}}$ is the vector of
Pauli matrices.

\section{Perfect localization due to interference effects in loops}

We begin considering the square diamond loop of fig.1b). The dots A and B
represent the input and the output leads, respectively. In the following we
neglect backscattering effects at the contacts assuming that the electrons
enter A with probability $1/2$ in the clockwise path AB and with probability 
$1/2$ in the counterclockwise path. The transmission amplitudes matrix ${\bf %
\Gamma }$ in B is 
\[
{\bf \Gamma }={\bf t}\,\,e^{i\,2\xi L} 
\]
where $t$ is the spin transmission matrix 
\[
{\bf t}=\left| 
\begin{array}{cc}
t_{\uparrow \uparrow } & t_{\uparrow \downarrow } \\ 
t_{\downarrow \uparrow }^{} & t_{\downarrow \downarrow }
\end{array}
\right| 
\]
given by the interference between the different spin precessions along the
two paths: 
\[
{\bf t}=\frac 12\left( {\bf R}_{SO}\left( -\,\,\frac \pi 4\right) {\bf R}%
_{SO}\left( \frac \pi 4\right) +{\bf R}_{SO}\left( \frac \pi 4\right) {\bf R}%
_{SO}\left( -\,\,\frac \pi 4\right) \right) . 
\]
It is simple to show that 
\begin{eqnarray*}
t_{\downarrow \downarrow } &=&t_{\uparrow \uparrow }^{*}=\frac 12(\cos
2k_0L+\cos 2k_1L+i\sqrt{2}\sin 2k_0L) \\
t_{\uparrow \downarrow } &=&t_{\downarrow \uparrow }^{}=\frac i{\sqrt{2}}%
\sin 2k_1L\;.
\end{eqnarray*}
Without the Dresselhaus term ($k_1=0$) the off diagonal elements of ${\bf t}$
matrix vanish and the spin up and spin down states do not interfere.
Assuming that the input is an unpolarized statistical mixture 
\[
\rho _{in}=\frac 12\left( \left| \uparrow \rangle \langle \uparrow \right|
+\left| \downarrow \rangle \langle \downarrow \right| \right) 
\]
the output will be described by\cite{Marigliano} 
\[
\rho _{out}=\frac 12\left( T_{\uparrow }\left| 1\rangle \langle 1\right|
+T_{\downarrow }\left| 2\rangle \langle 2\right| \right) , 
\]
where $T_{\uparrow }=\left| t_{\uparrow \uparrow }\right| ^2+\left|
t_{\downarrow \uparrow }^{}\right| ^2$ is the coefficient transmission for
an incoming spin up state and $T_{\downarrow }=\left| t_{\uparrow \downarrow
}\right| ^2+\left| t_{\downarrow \downarrow }\right| ^2$ is that for an
incoming spin down state. The spinors in $\rho _{out}$ are 
\[
|1\rangle =\frac 1{\sqrt{T_{\uparrow }}}\left( 
\begin{array}{c}
t_{\uparrow \uparrow } \\ 
t_{\downarrow \uparrow }^{}
\end{array}
\right) \,\,\,\,\,\text{ and \thinspace \thinspace \thinspace \thinspace
\thinspace \thinspace \thinspace }|2\rangle =\frac 1{\sqrt{T_{\uparrow }}}%
\left( 
\begin{array}{c}
t_{\uparrow \downarrow } \\ 
t_{\downarrow \downarrow }
\end{array}
\right) 
\]
corresponding to input spin up and down, respectively. Finally the
transmission coefficient of the unpolarized electrons is 
\begin{equation}
T=\frac 12\left( T_{\uparrow }+T_{\downarrow }\right) =\frac 12\left( \left|
t_{\uparrow \uparrow }\right| ^2+\left| t_{\downarrow \downarrow }\right|
^2+\left| t_{\uparrow \downarrow }\right| ^2+\left| t_{\downarrow \uparrow
}^{}\right| ^2\right) =  \label{eq10}
\end{equation}
\[
=\frac 14(\cos 2k_0L+\cos 2k_1L)^2+\frac 12(\sin ^22k_0L+\sin ^22k_1L). 
\]
Neglecting the Dresselhaus term ($k_1L=0$), eq.(\ref{eq10}) provides the
known result 
\begin{equation}
T=\cos ^2k_0L\left( 1+\sin ^2k_0L\right)  \label{eq11}
\end{equation}
that gives perfect localization ($T=0$) when $k_0L=n\pi /2\,\,\,(n=1,2,...)$%
\cite{Bercioux3}. When we begin to add gradually the Dresselhaus term, the
perfect localization is lost and the zeroes of $T$ become transmission
minima. Increasing more and more the Dresselhaus SO strength the perfect
localization is recovered when $k_1L=\pi /2$ and a new set of $T=0$ points
is obtained corresponding to $k_0L=n\pi \,\,\,(n=0,1,2,...).$ As shown in
fig.1a) a further increase of $k_1L$ generates a regular lattice of $T$
zeroes in the ($k_0L,k_1L$) plane given by: 
\begin{eqnarray*}
k_0L &=&n\pi /2\,\,\,\,(n=1,2,..),\,\,\,\,\,k_1L=(m-1)\pi \text{\thinspace
\thinspace \thinspace \thinspace }(m=1,2,..) \\
k_0L &=&(m-1)\pi \,\,\,\,\text{\thinspace \thinspace }(m=1,2,..),\,\,\,\,%
\,k_1L=n\pi /2\text{\thinspace \thinspace \thinspace \thinspace }(n=1,2,..).
\end{eqnarray*}

This result shows that we can get perfect localization in the diamond loop
of fig.1b with both the spin-orbit couplings. On the other hand we stress
that the foregoing result depends strictly on the angular position of the
loop with respect to the crystallographic axes of the substrate. Indeed the
geometry studied is somehow special. In order to consider a more general
case we analyse the same square loop rotated by an angle $\varphi $ with
respect to $x-$direction (see the inset of fig.2a)). The contour plots of $T$
as a function of $\varphi $ and of $k_0L$ are given in fig.2 for $k_1L=\pi
/4 $ and $\pi /2.$ For $k_1L=\pi /4$ there is no evidence of $T=0$ points at
any $\varphi .$ As the fig.2a) shows, only transmission minima are present
in this case. When $k_1L=\pi /2$ the zeroes of $T$ appear at $\varphi =\pi
/4,3\pi /4$ which corresponds to align the diagonal of the square loop along
the $x-$direction (fig.1b)). This results confirm that we get perfect
localization only for the pair ($k_0L$,$k_1L)$ shown in fig.1a): tilting the
square the zeroes transform in minima.

In order to make our analysis more complete, we considered also the
polygonal loops shown in the insets of fig.3: a rhombus and a six sided
cell. For the rhombus 
\[
{\bf t}=\frac 12\left( {\bf R}_{SO}\left( 0\right) {\bf R}_{SO}\left( \theta
\right) +{\bf R}_{SO}\left( \theta \right) {\bf R}_{SO}\left( 0\right)
\right) 
\]
while for the exagonal loop we get 
\[
{\bf t}=\frac 12\left( {\bf R}_{SO}\left( \theta \right) {\bf R}_{SO}\left(
0\right) {\bf R}_{SO}\left( -\theta \right) +{\bf R}_{SO}\left( -\theta
\right) {\bf R}_{SO}\left( 0\right) {\bf R}_{SO}\theta \right) . 
\]
From these transmission matrices the transmission coefficient for
unpolarized electrons can be obtained as we have shown in eq.(\ref{eq10}). A
careful analysis shows that specific values of $\theta $ exist such that,
again, we get the perfect localization ($T=0$). For such values the
vanishing of the transmission appears at some particular pairs of values ($%
k_1L,k_0L$) that are not connected continously with the $k_1=0$ zeroes. In
table 1 we report the values of $k_1/k_0$, $\theta $ and $k_0L$
corresponding to a perfect localization $T=0$ for unpolarized electrons. The
fig.3 reports contour plots of the transmission as a function of $\theta $
and $k_0L$ at the indicated vaues of $k_1L$. The zeroes of $T$ appear as
particular points at some specific values of the angle $\theta $ and of the
spin-orbit strengths. A regular pattern of zeroes is a special feature of
the square loop configuration of fig.1a) and it is lost for other polygonal
loop's shapes. 
\begin{eqnarray*}
&& 
\begin{array}{ccc}
k_1/k_0 & \theta & k_0L \\ 
0.3126 & 2.0885 & 10.4949 \\ 
0.2655 & 2.0313 & 13.6636 \\ 
0.2126 & 1.9572 & 19.9739 \\ 
0.5015 & 2.2464 & 21.0616 \\ 
0.3971 & 2.1721 & 21.8987 \\ 
0.2986 & 2.0723 & 22.5772 \\ 
0.3620 & 2.1402 & 25.1089 \\ 
0.2754 & 2.0439 & 25.7455 \\ 
0.4130 & 2.1853 & 27.5851
\end{array}
\,\,\,\,\,\,\, 
\begin{tabular}{lll}
$k_1/k_0$ & $\;\;\,\theta $ & $k_0L$ \\ 
0.4996 & 0.7896 & 3.0048 \\ 
0.2500 & 0.7879 & 6.2055 \\ 
0.1667 & 0.7867 & 9.3715 \\ 
0.2968 & 1.0192 & 10.2090 \\ 
0.3749 & 0.8207 & 12.2376 \\ 
0.1250 & 0.7862 & 12.5262 \\ 
&  &  \\ 
&  &  \\ 
&  & 
\end{tabular}
\\
&&\text{ \thinspace }\,\,\,\,\,\,\,\,\,\,\,\,\,\,\,\,\text{a) Rhombus
\thinspace \thinspace \thinspace \thinspace \thinspace \thinspace \thinspace
\thinspace \thinspace \thinspace \thinspace \thinspace \thinspace \thinspace
\thinspace \thinspace \thinspace \thinspace \thinspace \thinspace \thinspace
\thinspace \thinspace \thinspace \thinspace \thinspace \thinspace b)
Exagonal loop} \\
&&\,\,\,\,\,\,\,\,\,\,\,\,\,\,\,\,\,\,\,\,\,\,\,\,\,\,\,\,\,\,\,\,\,\,\,\,\,%
\,\,\,\,\,\,\,\,\,\,\,\,\,\,\,\,\,\,\,\,\text{Table 1}
\end{eqnarray*}

\section{Regulating the Al'tschuler--Aronov--Spivak and the Aharonov--Bohm
oscillations by means of Dresselhaus coupling}

In this section we discuss the effect of an external magnetic field $B$ on
the transmission properties of a 1D loop under both Rashba and Dresselhaus
interactions. We consider, first, a rhombic loop where the injection and the
collection nodes coincide with the A node in the inset of fig.3a (AA
configuration). In other words we are supposing that there are two possible
outputs at the collecting point, allowing the oscillation of the signal This
geometry has recently proposed by Koga et al.\cite{Koga2} to obtain a
ballistic spin interferometer where the collecting point is a splitter in
both incoming and outgoing directions. They use the cancelation of the AAS
oscillations due to Rashba SO, in the square loop shown in the inset of
fig.4a, to achieve an interferometric measure of SO strength $k_0.%
\mathop{\rm Si}%
$nce, as we will show in eq.(\ref{eq12}), the transmission coefficient in
presence of a magnetic field can be written in terms of that at zero
magnetic field, we start to discuss the latter case. In the AA configuration
the transmission amplitude matrix at zero magnetic field stems out from the
interference between the clockwise (CW) and the counterclockwise (CCW) paths
as 
\[
{\bf \Gamma }=\frac 12({\bf R}_{SO}\left( x,-\pi ,r\right) \cdot {\bf R}%
_{SO}\left( x,-\pi +\theta ,r\right) \cdot {\bf R}_{SO}\left( x,0,r\right)
\cdot {\bf R}_{SO}\left( x,\theta ,r\right) 
\]
\begin{eqnarray*}
&&\left. +{\bf R}_{SO}\left( x,\theta -\pi ,r\right) \cdot {\bf R}%
_{SO}\left( x,-\pi ,r\right) \cdot {\bf R}_{SO}\left( x,\theta ,r\right)
\cdot {\bf R}_{SO}\left( x,0,r\right) \right) e^{i\xi 4L} \\
&=&t_0\left( x,\theta ,r\right) \,\,e^{i\xi 4L}\cdot {\bf 1}
\end{eqnarray*}
where 
\[
{\bf R}_{SO}\left( x,\theta ,r\right) =\cos xy\otimes {\bf 1-}i\sin xy\left(
\sin 2\nu _{+}\otimes {\bf \sigma }_x+\cos 2\nu _{+}\otimes {\bf \sigma }%
_z\right) 
\]
and 
\[
x=k_0L,\,y\left( \theta ,r\right) =\sqrt{1+r^2+2r\sin 2\theta }=k_{so}\left(
\theta \right) /k_0\,\,\,
\]
with 
\[
r=k_1/k_0\,\,\,\,\,\,\text{and\thinspace \thinspace \thinspace \thinspace
\thinspace }\nu _{+}=\arctan \frac{\cos \theta +r\sin \theta -y}{\sin \theta
+r\cos \theta }.
\]
It is worth to note that the input spin state is conserved and the
transmission coefficient 
\[
T_0\left( x,\theta ,r\right) =t_0^2\left( x,\theta ,r\right) 
\]
is plotted in fig.4a for $\theta =\pi /2$ and in fig.5a for $\theta =\pi /4.$

In presence of a magnetic flux the matrix of the transmitted amplitudes is
no longer diagonal and becomes: 
\[
{\bf \Gamma }=\frac 12\left[ {\bf R}_{SO}\left( x,-\pi ,r\right) \cdot {\bf R%
}_{SO}\left( x,-\pi +\theta ,r\right) \cdot {\bf R}_{SO}\left( x,0,r\right)
\cdot {\bf R}_{SO}\left( x,\theta ,r\right) e^{i\phi /2}\right. + 
\]
\begin{equation}
\left. {\bf R}_{SO}\left( x,\theta -\pi ,r\right) \cdot {\bf R}_{SO}\left(
x,-\pi ,r\right) \cdot {\bf R}_{SO}\left( x,\theta ,r\right) \cdot {\bf R}%
_{SO}\left( x,0,r\right) e^{-i\phi /2}\right] e^{i\xi 4L}  \label{eq12a}
\end{equation}
\[
=\left| 
\begin{array}{cc}
t_{\uparrow \uparrow }\left( x,\theta ,r,\phi \right) & t_{\uparrow
\downarrow }^{}\left( x,\theta ,r,\phi \right) \\ 
t_{\downarrow \uparrow }^{}\left( x,\theta ,r,\phi \right) & t_{\downarrow
\downarrow }^{}\left( x,\theta ,r,\phi \right)
\end{array}
\right| e^{i\xi 4L} 
\]
with 
\begin{eqnarray*}
t_{\uparrow \uparrow } &\neq &t_{\downarrow \downarrow }^{} \\
t_{\downarrow \uparrow }^{} &=&t_{\uparrow \downarrow }^{*}\neq 0.
\end{eqnarray*}
In eq.(\ref{eq12a}) the rhombus (with area $S$) is threaded by a magnetic
flux $\Phi =BS=\phi \Phi _0$ where $\Phi _0=h/2e$ is the magnetic flux half
quanta. The input spin state is no more conserved: the interference between
CW and CCW paths is able to rotate the spin. The transmission coefficient
for unpolarized electrons can be, then, written as 
\begin{equation}
T\left( x,\theta ,r,\phi \right) =\frac 12\left( \left| t_{\uparrow \uparrow
}\right| ^2+\left| t_{\downarrow \uparrow }^{}\right| ^2+\left| t_{\uparrow
\downarrow }^{}\right| ^2+\left| t_{\downarrow \downarrow }^{}\right|
^2\right) =\frac 12+\left( T_0\left( x,\theta ,r\right) -\frac 12\right)
\cos \phi .  \label{eq12}
\end{equation}
As already mentioned, the AAS oscillations are given by the term $\cos \phi $
whose prefactor contains the zero field transmission $T_0$, that is all we
need to perform the analysis of the magnetic field effects. For a square
loop ($\theta =\pi /2$) and without the Dresselhaus term ($k_1L=0$) we
recover the known result by Koga et al.\cite{Koga2} 
\[
T_0\left( x,\frac \pi 2,0\right) =\left( \cos ^2x+\cos 2x\sin ^2x\right) ^2 
\]
that is plotted in fig.4a (dashed curve). The perfect localization ($T=0\,$)
is obtained when $x=\pi /2,\pi $ at $\phi =\pi .$ Eq.(\ref{eq12}) shows that
when $T_0=1/2$ the AAS oscillations are suppressed. On the other hand the
transmission $T$ assumes the same costant value $1/2$ when $\phi =\pi
/2,\,3\pi /2$ and, at these magnetic fluxes, the modulation of the
transmission due to SO couplings is cancelled. Koga, Sekine and Nitta\cite
{Koga3} have realized experimentally a Rashba ballistic spin interferometer
using a network of square loops. They measured the conductivity $s$ varying
the magnetic field and controlling the strength of the Rashba term by means
of a gate voltage. Assuming that the conductivity, in the ballistic regime,
is proportional to the transmission coefficient (\ref{eq12}). They searched
the values of \thinspace \thinspace $x$ for which $s$ becomes independent on
the magnetic field $B$ in a range around $B=0$, and from these values they
obtained a measure of Rashba SO strength $k_0$.

The zero field transmission when also the Dresselhaus term is added (for the
square loop) is shown in fig.4. Also in this case the AAS oscillation are
suppressed at the $x_{*}\left( r\right) $ values for which 
\[
T_0\left( x_{*},\theta ,r\right) =\frac 12. 
\]
The fig.4b shows the values of $x_{*}=k_0L$ at which the suppression of AAS
is obtained as a function of the ratio between the Dresselhaus and Rashba
strength, $r$. Increasing $r$ the period of $T_0$ decreases from the value $%
\pi $ at $r=0$ to lower values. The two zeroes of $T$ approach each other
and disappear at $x=1.451$ for $r=0.414213$. For $k_1=k_0,$ $T_0(x,\pi
/2,1)=1$ and the transmission coefficient becomes independent of the
spin-orbit coupling. The fig.4b shows that for $r<0.199$ we have four AAS
suppression points that become two when $0.199<r<0.668.$ The cancellation of
AAS oscillations is not possible for greater values of Dresselhaus strength (%
$r>0.668$). This analysis shows how relevant the inclusion of Dresselhaus
term is in order to describe in a proper way the AAS suppression.
Furthermore our study allows an extension of the ballistic spin
interferometric technique developed by Koga et al. \cite{Koga2} that could
be used also to measure the ratio between the Rashba and Dresselhaus terms.

To investigate if the AAS suppression depends on the shape of the
interferometer we have taken into account a different rhombus geometry with $%
\theta =\pi /4.$ The fig.5a shows the transmission at zero field, and the
fig.5b shows how the suppression points change with $r$. The cancellation of
AAS oscillation is still present though the pairs of values at which AAS
suppression occurs $(k_{0,}k_1)$ change modifying the shape. The supression
remains also when the loop is rotated with respect to the substrate.

To conclude the analysis of the magnetic field effects let us consider what
happens if the electrons are injected in the node A and collected in the
opposite node B, traversing the square loop (AB configuration). In this case
the transmission amplitudes matrix is given by 
\begin{eqnarray*}
{\bf \Gamma } &=&\frac 12({\bf R}_{SO}\left( x,\pi /2,r\right) \cdot {\bf R}%
_{SO}\left( x,0,r\right) e^{i\phi /4}+{\bf R}_{SO}\left( x,0,r\right) \cdot 
{\bf R}_{SO}\left( x,\pi /2,r\right) e^{-i\phi /4})e^{i2\xi L} \\
&=&\left| 
\begin{array}{cc}
t_{B\uparrow \uparrow } & t_{B\uparrow \downarrow } \\ 
t_{B\downarrow \uparrow }^{} & t_{B\downarrow \downarrow }
\end{array}
\right| e^{i2\xi L}.
\end{eqnarray*}
For unpolarized electrons the transmission coefficient becomes 
\[
T_B\left( x,r,\phi \right) =\frac 12\left( \left| t_{B\uparrow \uparrow
}\right| ^2+\left| t_{B\downarrow \downarrow }\right| ^2+\left| t_{B\uparrow
\downarrow }\right| ^2+\left| t_{B\downarrow \uparrow }^{}\right| ^2\right) =%
\frac 12+\left( T_B\left( x,r,0\right) -\frac 12\right) \cos \frac \phi 2 
\]
The factor $\cos \left( \phi /2\right) $ describes the Aharonov-Bohm
oscillations\cite{AB}, which present a double period with respect to the AAS
oscillations, and, again, his prefactor is fixed by the zero field
transmission $T_B\left( x,r,0\right) $ that regulates the amplitude of AB
oscillation. This quantity is plotted in fig.6a. As for the foregoing AA
configuration $T_B\left( x,r,0\right) =1/2$ implies that $T_B\left( x,r,\phi
\right) =1/2$ for any $\phi $ and the ratio $r=k_1/k_0$ can be fixed in such
a way that the AB oscillations are cancelled. Therefore, the suppression
takes place at $x$ values satisfying the equation 
\[
T_B\left( x_{AB}(r),r,0\right) =\frac 12. 
\]
The behaviour of the AB square configuration is shown in fig.6.

\section{Concluding remarks}

In conclusion we have studied the interference effects due to the Rashba and
the Dresselhaus SO interactions in quantum wires forming polygonal loops.
The spin precession along the sides of the loop gives rise to perfect
localization at particular values of the pair ($k_1L$ , $k_0L$). For the
square diamond loop we achieve the perfect localization for pairs $%
(k_0L,\,k_1L)$ belonging to a square lattice that is symmetrical with
respect to the two SO strengths $k_0$ and $k_1$. The periodic pattern of the
transmission zeroes\cite{Bercioux3} obtained with only the Rashba SO
interaction\cite{Bercioux3}, is preserved adding Dresselhaus SO coupling.
The configuration with the square diagonal parallel to $x-$axis (in [010]
crystallographic direction) is a special case and when the square is rotated
in $x-z$ plane the zeroes of $T$ transform in minima and the perfect
localization is lost. We have studied other two geometries: a rhombus and an
exagonal cell. For both cases pairs $(k_0L,\,\,k_1L)$ exist that give the
perfect localization only for a specific shape (we characterize the shape
with an angular opening $\theta $). We have found triplets $(\theta
,k_0L,k_1L)$ that give transmission zeroes. This behaviour suggests that the
perfect localization in a circular loop is not easy to predict. In
particular, the procedure discussed in ref.\cite{Bercioux3} in the case of
Rashba coupling, where perfect localization in a circle is obtained as a
limit of a succession of regular polygons, cannot be applied in the same
way. The perfect localization on a circle with both the SO couplings will be
matter of future research.

When the loop is plunged in an external magnetic field the transmission
coefficient oscillates with the magnetic flux passing through the loop. The
amplitude of this oscillation depends on the strengths of the two SO
couplings. Injecting and collecting the electrons at the same loop node (the
interfering paths are self-itersecting ones), the 1D loop behaves as a
ballistic spin interferometer. With this configuration the AAS oscillations
appear and, in presence of Rashba SO, they are suppressed for some
particular values of $k_0L$\cite{Koga2}. We have considered an
interferometer with the shape of a rhombus with both the SO interactions.
The suppression appears at $k_0L$ values which depend on the ratio $k_1/k_0$%
. So that the interferometric experimental technique of Koga et al.\cite
{Koga3} could be used to measure not only the $k_0$ value but also the ratio 
$k_1/k_0$. An other kind of magnetic modulation of the transmission
coefficient are the AB oscillations whose period is the double of the AAS
oscillations. They appear when the electrons are injected and collected at
opposite nodes of the loop and the interfering paths of equal length surrond
the loop area. Again the presence of the Dresselhaus coupling can regulate
the amplitude of these oscillations.

Our results concern a single loop. When the loops are arranged in a quantum
network the transport properties through the system may change as discussed,
for the Rashba SO case in Refs. \cite{Bercioux1,Bercioux2}. We also expect
that the use of more realistic boundary conditions could be important, for
example the finite coupling with leads can give resonances representing
quasibound states within the loop.

To conclude we briefly discuss the consequences of higher order winding
contributions and backscattering. The simplest way to deal with this
question is to combine the multiple scattering against the injection node
and the collector node incoherently\cite{Dattabook}. Then the single
scattering event can be characterized with a classical probability. We
identify the probability that the electron leaves a node with the
transmission coefficients $T$ that we have calculated before, the classical
reflection probability being $R=1-T$. The round trips can be arranged into a
geometrical series\cite{Dattabook} whose sum gives the composite exit
probability ${\cal T}$%
\[
{\cal T=}\frac{T^2}{1-R^2}=\frac T{2-T}\;. 
\]

We note that $T=0,1$ implies that also ${\cal T}=0,1$. The total
transmission ${\cal T}$ \/ keeps the periodicity in $\phi $ although the
dependence on $\phi $ is no more simply $\cos \phi $ or $\cos 2\phi $ as
before. Therefore, this assumption of incoherence predicts that the perfect
localization and the suppression of AAS and AB oscillations are not spoiled
by incoherent multiple scattering. The transmission ${\cal T}$ \/ becomes
independent on $\phi $ $\,$at some particular values $k_0L$ in the same way
as $T,$ with the same dependence on the ratio $r=k_1/k_0$, but the value of $%
{\cal T}$ \/at the suppression lowers from $1/2$ to $1/3$.

\begin{center}
{\bf ACKNOWLEDGMENTS}
\end{center}

We acknowledge Dario Bercioux for a critical reading of the manuscript. We
warmly thank Diego Frustaglia for useful suggestions to clarify some point
of the paper.

\begin{center}
{\bf FIGURE\ CAPTIONS}
\end{center}

\begin{enumerate}
\item[Fig.1]  Perfect localization in the diamond square loop. In a) there
is the countour plot of the transmission as a function of $k_0L$ and of the
ratio $k_1/k_0$. The part b) shows the zeroes of $T$ in $k_0L$, $k_1L$
plane. In c) is shown the square with the diagonal parallel to $x-$axis for
which the perfect localization occurs.

\item[Fig.2]  Contour plots of the transmission coefficient of the rotated
square diamond loop as a function of $k_0L$ and of the rotation angle $%
\varphi $ at the two indicated values of $k_1L$. The zeroes of $T$ appear
only for $\varphi =\pi /4,3\pi /4$ for $k_1L=\pi /2.$

\item[Fig.3]  Contour plots of the transmission coefficient of the rhombus
and of the exagonal cell as a function of $k_0L$ and of the angular opening $%
\theta $, at the indicated values of $k_1L$ at which an isolated zero of $T$
appear.

\item[Fig.4]  a) Transmission coefficient of the square loop $T_0$ at zero
magnetic field for the square ($\theta =\pi /2$) interferometer (electrons
enter an exit in A) versus $k_0L$ at the indicated values of $r=k_1/k_0$.

b) Plot of the value $x_{*}$ of $k_0L$ as a function of $r$ for which $%
T_0=1/2$ an the the the AAS oscillations are suppressed.

\item[Fig.5]  The same plots of fig.4 for a rhombus with angular opening $%
\theta $ of $\pi /4$

\item[Fig.6]  The suppression of the Aharonov--Bohm oscillations in the
square loop (the electrons enter in A and are collected in B). The solutions 
$x\left( r\right) $ of the equation $T_{AB}\left( x\left( r\right)
,r,0\right) =1/2$ are shown in the part b).
\end{enumerate}


\begin{references}
\bibitem{nota1}  The limit of this approximation has been studied in refs.(%
\cite{Mireles},\cite{Zulicke})
\bibitem{nota2}  The same result for $T$ is obtained including only the
Dresselhaus term ($k_0=0$) and replacing in eq.(\ref{eq11}) $k_0$ with $k_1$.
\end{references}
\end{document}